\documentclass{article}

\usepackage[dvips]{graphicx}
\usepackage{amscd}
\usepackage{xspace}
\usepackage{amstext}
\usepackage{amssymb,amsfonts,amsthm}
\usepackage{amscd}
\usepackage{hyperref}
\usepackage{psfrag}
\newcommand{\p}{\partial}

\newcommand{\dd}{{\rm d}}

\newtheorem{theorem}{Theorem}[section]

\begin{document}

\title{Simultaneity in special and general relativity\footnote{To appear in the proceedings of
the school {\em Relativistic Coordinates, Reference and
Positioning Systems}, Salamanca (Spain), January 21-25, 2005 }}

\author{E. Minguzzi\\{\small Departamento de Matem\'aticas,}  {\small Universidad de Salamanca}, \\ {\small Plaza
de la Merced 1-4, E-37008 Salamanca, Spain} \\ {\small and INFN,
Piazza dei Caprettari 70, I-00186 Roma, Italy}
 \\ {\small minguzzi@usal.es} }

\date{}
\maketitle

\begin{abstract}
We present  some basic facts concerning simultaneity in both
special and general relativity. We discuss  Weyl's proof of the
consistence of Einstein's synchronization convention and consider
the general relativistic problem of assigning a time function to a
congruence of timelike curves.
\end{abstract}


\section{Simultaneity in special relativity}
\label{sec:1}

In special relativity the possibility of synchronizing distant
clocks so as to obtain a global coordinate time was proved by Weyl
\cite{weylG} in a fundamental and unfortunately overlooked proof.
Let a light beam travel from point $A$ to point $B$, leaving $A$
at time $t_A$ according to the clock  at $A$. The Einstein
convention states (in Weyl's version) that  the clock at $B$ (say
clock $B$ for short) is Einstein synchronized with clock $A$ if
the time of arrival according to clock $B$ reads
$t_B=t_A+\overline{AB}/c$ where $\overline{AB}$ is the Euclidean
distance between the two points. This synchronization convention
should satisfy at least the following three properties in order to
lead to a spacetime foliation into (equal time) simultaneity
slices
\begin{itemize}
\item[(i)] {\bf Time homogeneity}. If clock $B$ is set accordingly to the
procedure above then repeating the same experiment the equation
$t_B=t_A+\overline{AB}/c$  holds for whatever value of $t_A$
without the need of setting again
clock $B$ (i.e. it is meaningful to say that clock $B$ is synchronized with clock $A$). \\
\item[(ii)] {\bf Symmetry}. If  clock  $A$  is synchronized with clock $B$ then clock $B$ is
synchronized with clock $A$ (i.e.  it is meaningful to say that the two clocks {\em are} synchronized). \\
\item[(iii)] {\bf Transitivity}. If clocks $A$ and $B$ are synchronized and clocks $B$ and
$C$ are synchronized then clocks $A$ and $C$ are synchronized.
\end{itemize}
It could be tempting to consider the equation
$t_B=t_A+\overline{AB}/c$ as a trivial consequence of the special
relativistic postulate of the constancy of the speed of light.
However, here we should take into account that before making any
statement on the value of the {\em one-way} (i.e. from one point
to another) speed of light, and thus even before the formulation
of the constancy postulate, a global time variable to make sense
of expressions such as $\Delta x/\Delta t$ is needed. In most
special relativity textbook this important conceptual point is not
explained, and the existence of a global time variable such that
the one-way speed of light is a constant $c$ is assumed without
further explanations. This approach was initiated by Einstein  in
his 1905 work ``On the electrodynamics of moving bodies"
\cite{einstein05} where he writes
\begin{quotation}
\noindent Let a ray of light start at the ``A time'' $t_A$ from A
towards B, let it at the ``B time'' $t_B$  be reflected at B in
the direction of A, and arrive again at A at the ``A time''
$t'_A$.

In accordance with definition the two clocks synchronize
if\footnote{This definition of clock synchronization was first
given by Poincar\'e \cite{poincare04a,poincare04b} although at the
time  it was already applied by engineers in the calculation of
longitudes by means of telegraphic signals \cite{galison03}.}
\[
t_B-t_A=t'_A-t_B . \] We assume that this definition of
synchronism is free from contradictions, and possible for any
number of points; and that the following relations are universally
valid:--
\begin{enumerate}
\item If the clock at B synchronizes with the clock at A, the clock
at A synchronizes with the clock at B. \\
\item If the clock at A
synchronizes with the clock at B and also with the clock at C, the
clocks at B and C also synchronize with each other.
\end{enumerate}
Thus with the help of certain imaginary physical experiments we
have settled what is to be understood by synchronous stationary
clocks located at different places, and have evidently obtained a
definition of ``simultaneous,'' or ``synchronous,'' and of
``time.'' The ``time'' of an event is that which is given
simultaneously with the event by a stationary clock located at the
place of the event, this clock being synchronous, and indeed
synchronous for all time determinations, with a specified
stationary clock.

In agreement with experience we further assume the quantity
\[
\frac{2\overline{AB}}{t'_A-t_A}=c .
\]
to be a universal constant -- the velocity of light in empty
space.
\end{quotation}

Thus Einstein assumes without proof the validity of (ii) and
(iii). On the contrary, Weyl maintains that in a correct
conceptual development of the theory  a clock synchronization
convention should be defined and its coherence proved (i.e. its
symmetry and transitivity). Finally, it should be shown that with
respect to the global time variable provided by the clocks'
readings the one-way speed of light is $c$.

Weyl succeeded in completing this program starting from the
experimental fact that the speed of light is a constant $c$ around
any closed polygonal path (the light beam can be reflected over
suitable mirrors so as to travel over the closed polygon).
Contrary to the one-way speed of light, the average speed of light
over a closed path is independent of the distant synchronization
convention adopted  and as such the statement that the speed of
light over closed paths is a constant is indeed well defined prior
to the construction of a global time variable. Under the
reasonable assumption that the speed of light is a constant over
closed polygons (it was reasonable since it could be tested) Weyl
was able to prove (ii) and (iii) assuming (i) tacitly. The
assumption (i) was  recognized and removed in \cite{minguzzi02d}.
Since the complete proof is short we give it here; for more
details the reader is referred to \cite{minguzzi02d}

\begin{itemize}
\item[(i)] Emit a light beam at time $t_A$ from point $A$.
Let it arrive at $B$ where clock $B$ is set so that the time of
arrival is $t_B=t_A+\overline{AB}/c$. We have to prove that if a
second light beam is emitted at time $t'_A>t_A$ in direction of
$B$, the time of arrival at $B$ is given by the same equation
$t'_B=t'_A+\overline{AB}/c$.

Consider a point $C$ at a distance from $B$ given by
$\overline{BC}=\frac{c}{2}(t'_B-t_B)$ and assume that the second
light beams once arrived at $B$ is there reflected back to $A$
 that it reaches at time $t''_{A}$. Moreover, assume that the first
 light beam once reached $B$ is there reflected to $C$, then again to $B$ and finally $A$. We have three closed
polygonal paths of interest, $ABA$, $ABCBA$ and $BCB$. Since the
speed of light is $c$ over any closed polygonal path we have the
equations
\begin{eqnarray}
2\overline{AB}&=& c (t''_{A}-t'_A)\\
2\overline{AB}+2\overline{BC}&=&c(t''_{A}-t_A) \\
2\overline{BC}&=&c(t'_B-t_B)=c(t'_B-t_A)-\overline{AB}
\end{eqnarray}
Summing the first and the third equation and subtracting the
second we obtain the thesis.

\item[(ii)] Emit a light beam at time $t_A$ from point $A$.
Let is arrive at $B$ where clock $B$ is set so that the time of
arrival is $t_B=t_A+\overline{AB}/c$. We have to prove that if a
second light beam is emitted at time $t'_B$ in direction of $A$,
the time of arrival at $A$ is given by the same equation
$t'_A=t'_B+\overline{AB}/c$. Point (i) implies that the time
needed by the light beam to reach $A$ departing from $B$ is
independent of the instant of departure. We can therefore assume
without loss of generality that $t'_B=t_B$. Then we can consider
the two light beams as a single light beam that, reflected at $B$,
covers the closed path $ABA$. Thus
\begin{equation}
2\overline{AB}=c(t'_A-t_A)=c(t'_A-t_B)+\overline{AB}=c(t'_A-t'_B)+\overline{AB}
\end{equation}
from which the thesis follows.

\item[(iii)] Place two clocks at $A$ that we denote with clock $A$ and clock $A'$.
From (i) and (ii) it follows that it makes sense to say that two
distant clocks are synchronized. Consider  the three points $A$,
$B$ and $C$ with their respective clocks.  Assume that $A$ and $B$
are synchronized, and that $B$ and $C$  are synchronized. We have
to prove that $A$ and $C$ are synchronized. To this end
synchronize $A'$ with $C$ and send a light beam all over the
closed polygonal path $ABCA$. Let the sequence of time of
arrivals/departures be given by $t_A$, $t_B$, $t_C$, $t'_A$ and
denote with $t'_{A'}$ the time at which the light beam returns at
$A$ according to clock $A'$. We have from the constancy of the
speed of light over polygonal paths
\begin{equation}
c(t'_{A}-t_A)=\overline{AB}+\overline{BC}+\overline{AC}
\end{equation}
but since the clock pairs $A-B$, $B-C$ and $C-A'$ are synchronized
we have
\begin{eqnarray}
t_B=t_A+\overline{AB}/c\\
t_C=t_B+\overline{BC}/c\\
t'_{A'}=t_C+\overline{AC}/c\\
\end{eqnarray}
and summing the three equations we obtain
\begin{equation}
c(t'_{A'}-t_A)=\overline{AB}+\overline{BC}+\overline{AC}
\end{equation}
or $t'_{A'}=t'_{A}$, that is the measures of clock $A$ coincide
with those of clock $A'$ and therefore clock $A$ is synchronized
with clock $C$.
\end{itemize}
Once the Einstein synchronization convention is proved to be
coherent, from equation $t_B=t_A+\overline{AB}/c$ it follows that
the one-way speed of light is $c$ in the new global {\em Einstein}
time. From this fact one recovers that, coherently with the
hypothesis, the speed of light over closed paths is a constant
$c$.

Unfortunately in a rotating frame it can be easily shown using
special relativity that the speed of light over a closed path
depends on the direction followed and therefore can not be a
universal constant - this is the well known Sagnac effect
\cite{post67,ashtekar75}. Thus Weyl's theorem can not be applied
in a rotating frame and indeed the Einstein convention fails to be
transitive in this case.

A natural question is whether  a small correction $\delta(A,B)$
exists such that defining $t_B=t_A+\overline{AB}/c+\delta(A,B)$
the transitivity is restored. The answer is affirmative but it
requires some work to find its actual expression. In general
$\delta$ is expected to vanish in an inertial frame as the
Einstein convention works perfectly there. At the experimental
level it should therefore vanish if the vorticity and acceleration
of the frame vanish while it can be different from zero if these
quantities do not vanish.

\section{Simultaneity in curved spacetimes and non-inertial frames}
We recall that a frame in special and general relativity is a
congruence of timelike curves each one defining the motion of a
point ``at rest" in the frame. For instance, this paper is made of
points that in spacetime are represented by a worldline. Each
distinct point is represented by a different worldline and hence,
in general, the motion of a body on spacetime is represented by a
congruence of timelike curves defining the spacetime motion of its
points. Mathematically there will be a projection $\pi: M \to S$
from the spacetime manifold $M$ to the space $S$ that to each
event $m$ associates the worldline (space point) $s=\pi(m)$
passing through it. Locally one has the structure of a fiber
bundle where the fiber is diffeomorphic to $\mathbb{R}$. However
we are not in a principle fiber bundle since no action of the
group $(\mathbb{R},+)$ on $M$ has been defined.

This mathematical construction is known as the hydrodynamical
formalism of general relativity as the motion of the body is
regarded in a way analogous to the motion of a fluid in Euclidean
space \cite{hawking73}. We shall try to associate to the timelike
flow a foliation of spacetime in spacelike simultaneity slices.
Our concept of simultaneity will be therefore related to a
timelike flow, which can be regarded as the motion of a set of
observers in spacetime \cite{gao98}. Other authors consider
instead simultaneity foliations associated to a privileged
observer \cite{bolos02} (i.e. a privileged timelike worldline).

It is useful to choose coordinates on $M$ as follows. First
introduce space coordinates $\{ x^i\}$ on $S$ so that each space
point corresponds to a different triplet $x^i$, and then complete
the coordinate system with a time coordinate $t$, $\dd t \wedge
\dd x^1 \wedge \dd x^2 \wedge \dd x^3\ne 0$, $\p_t$ timelike.
There are many different ways in which $t$ can be introduced and
the surfaces $t=cnst.$ will be called simultaneity slices for the
given time function choice.

Unfortunately, this way of defining a global time variable is not
physically satisfactory since the time function is simply assumed
to exist without giving a constructive procedure or a way to
measure it. Moreover, it does not help in selecting a useful time
function or simultaneity convention. For instance the Einstein
time would be only one among many possible choices in Minkowski
spacetime.

We propose, therefore, to construct the global time function from
a local definition of simultaneity. A local definition of
simultaneity is an assignment to each spacetime event of a
spacelike hyperplane that, roughly speaking, determines the events
that are locally simultaneous. These hyperplanes are called
horizontal hyperplanes, and in the language of gauge theories they
define a connection: the {\em simultaneity connection}
\cite{minguzzi03}. If this connection is integrable, i.e. its
curvature vanishes, the distribution of horizontal hyperplanes is
integrable and gives rise to a spacetime foliation through
spacelike hypersurfaces: the hypersurfaces of simultaneity. For
instance, for the Einstein convention the horizontal hyperplanes
are those perpendicular to the 4-velocity field $u(x)$ of the
timelike flow and the curvature of the connection is proportional
to the vorticity of the flow.

By itself this constructive choice does not allow to reduce the
large arbitrariness already seen in the previous approach through
time functions. On the contrary it seems to make it even worse
since each time function $t(x)$ has hypersurfaces $t=cnst.$ whose
tangent hyperplanes determine  a connection. However, we now
require each allowed simultaneity connection to be a convention,
i.e. the distribution of horizontal hyperplanes should depend on
the spacetime point only through local measurable quantities
related to the spacetime structure and frame. Examples of such
quantities are the vorticity
$w^{\eta}=\frac{1}{2}\varepsilon^{\eta \beta \alpha \gamma}
u_{\beta} u_{\alpha;\gamma}$, acceleration
$a_{\mu}=u_{\mu;\alpha}u^{\alpha}$, expansion and shear of the
frame, and the metric or the curvature tensors.

This requirement implies that the observers can determine which
events are simultaneous according to the simultaneity convention
without the need of global information. For instance, in the
Einstein convention one synchronizes its own clock with those of
the few observers in its neighborhood: the observer does not need
to be in contact with distant observers. Despite the local nature
of the procedure the observer knows that the synchronization
convention is coherent and that for this reason it will provide a
global time variable.

At the mathematical level, let $u(x)$ be the normalized ($u^{\mu}
u_{\mu}=1$) 4-velocity field of the congruence of timelike curves.
The distribution of horizontal hyperplanes $H_u(x)$ is uniquely
determined as the Ker of a 1-form $\omega$ normalized so that
$\omega(u)=1$ and such that $\omega_{\mu}$ is timelike (otherwise
the horizontal hyperplane would not be spacelike). It is
convenient to introduce the vector product between the vorticity
vector and the acceleration, $m_{\alpha}=\epsilon_{ \alpha \beta
\gamma \delta} a^{\beta} u^{\gamma} w^{\delta}$ and limit  our
analysis only to those spacetime regions where $m_{\alpha}\ne 0$.
We also define $a^2=-a^{\mu} a_{\mu}$, $w^2=-w^{\mu}w_{\mu}$ and
$m^{2}=-m^{\mu}m_{\mu}=a^{2}w^{2}\sin^{2}\theta$ where $\theta$ is
the angle between the vorticity vector and the acceleration for an
observer moving at speed $u$. Since $u^{\mu}$, $a^{\mu}$,
$w^{\mu}$ and $m^{\mu}$ are linearly independent any local
simultaneity convention takes the form
\begin{equation}\label{generic}
\omega_{\alpha}=u_{\alpha}+\psi^{m}(x) m_{\alpha}+\psi^{a}(x)
a_{\alpha}+ \psi^{w}(x) w_{\alpha} ,
\end{equation}
for suitable functions $\psi^{m},\psi^{a},\psi^{w}$. From the
definition of local simultaneity convention it follows that
$\psi^{m},\psi^{a},\psi^{w}$, depend on the spacetime event $x$
through the acceleration $a$, the vorticity $w$, the angle
$\theta$ between them, and possibly on other scalars. Note that if
the $\psi$ functions are small the simultaneity connection may be
considered as a perturbation of Einstein's for which we have
$\omega_{\alpha}=u_{\alpha}$.

The remaining problem is that of finding a suitable local
simultaneity convention that reduces to Einstein's in the
Minkowskian-inertial frame case.
 We make some simplifying assumptions
\begin{itemize}
\item[(a)] The frame is generated by a Killing vector field $k$.
\item[(b)] The functions $\psi^{m}$, $\psi^{w}$ and $\psi^{a}$
are constructed from the observable quantities $a$, $w$ and
$\theta$ (or equivalently  $a$, $w$ and $m$ with $m=a w \sin
\theta$).
\item[(c)] The curvature of the 1-form connection $\omega_{\mu}$ is proportional
to the Riemann tensor (through contraction with a suitable
tensor).
\end{itemize}
The first two conditions are natural simplifications that allow us
to tackle the problem while keeping the calculations at a
reasonable size. The last one is imposed since the requirement
that the curvature of the corresponding gauge theory vanishes
would be too restrictive and no simultaneity connection satisfying
that requirement would be eventually found. With our condition
(c), at least in the weak field limit, the distribution of
horizontal planes becomes integrable providing a useful definition
of simultaneity. In particular it becomes exactly integrable in
Minkowski spacetime where the Riemann tensor vanishes.

The following theorem holds \cite{minguzzi04} \\
\begin{theorem}
In a stationary spacetime let $k$ be a timelike Killing vector
field and set $u=k/\sqrt{k \cdot k}$. Let $U$ be the open set
$U=\{x: m(x)
> 0 \ \textrm{and} \ a(x) \ne w(x) \}$. Consider in $U$ the connection
\begin{equation} \label{conn}
\omega_{\alpha}=u_{\alpha}+\psi^{m}(x) m_{\alpha}+\psi^{a}(x)
a_{\alpha}+ \psi^{w}(x) w_{\alpha}.
\end{equation}
Let $\psi^{m},\psi^{a},\psi^{w}$, be $C^{1}$ functions dependent
only on $a$, $w$ and $\theta$.  Then, regardless of the stationary
spacetime considered, the connection is timelike in $U$ (and hence
it is a simultaneity connection in $U$) and has a curvature
proportional to the Riemann tensor in $U$  only if
\begin{equation} \label{magic}
\psi^{m}=\frac{a^{2}+w^{2}-\sqrt{(a^{2}+w^{2})^{2}-4m^{2}}}{2m^{2}}. \\
\end{equation}
\end{theorem}
\noindent The theorem selects the simultaneity connection
\begin{equation} \label{cbar}
\bar{\omega}_{\alpha}=u_{\alpha}+
\frac{a^{2}+w^{2}-\sqrt{(a^{2}+w^{2})^{2}-4m^{2}}}{2m^{2}}\,
m_{\alpha},
\end{equation}
that we call $\bar{C}$-simultaneity, as the most natural and
useful in the week field limit. It seems remarkable that it
differs from the Einstein's simultaneity connection $\omega=u$.
The relation of $\bar\omega$ with the function $\bar\delta$ used
in practice can be found in \cite{minguzzi04}. The
$\bar{C}$-simultaneity connection proves particularly useful in
Minkowski spacetime since there it is exactly integrable. Contrary
to the Einstein convention that does not provide an integrable
foliation for observers in a rotating platform, the
$\bar{C}$-simultaneity connection proves to be integrable in that
case. The simultaneity slices turn out to be the same of the
inertial observers at rest in the inertial frame i.e. those that
observe the rotating platform from the outside. Thus, although the
rotating observers could in principle ignore to be at rest in a
rotating platform, the $\bar{C}$-simultaneity convention that they
apply allow them to define a natural global time function.

\section{Conclusions}
In the first section we have presented the proof of the
consistence of the Einstein synchronization convention starting
from the constancy of the speed of light over polygonal paths. As
far as we know, despite the relevance of this proof for any
conceptually rigorous development of special relativity, it could
not be found in any special relativity textbook. The only relevant
exception is the old 1923 book by Weyl which unfortunately has
never been translated from the German (the English translation
\cite{weylE} is of the 1921 German edition which does not contain
the proof; the relevant section on Einstein synchronization has
recently been translated in \cite{minguzzi02d}). We mention that
sometimes the proof is, in our opinion incorrectly, attributed to
Reichenbach. As a matter of fact Reichenbach introduced a
round-trip axiom - {\em the speed of light over a closed path is
the same in both directions} - which is weaker than Weyl's. Using
Reichenbach postulate it is not possible to prove the consistence
of Einstein's synchronization convention as Weyl did
\cite{minguzzi02d}. By the way, Weyl and Reichenbach were in
contact in the early twenties when their books on special
relativity appeared \cite{rynasiewicz05} so it is not surprising
that they considered similar postulates.

At the end of the first section we have pointed out that in many
circumstances, even in flat spacetime, the Weyl's theorem can not
be applied since the speed of light over closed paths is not
constant in rotating frames (the Sagnac effect). Therefore, in the
second section we have considered the problem of finding a
coordinate time in more general cases, i.e. in general relativity
and for non-inertial reference frames. We have introduced the
concept of local simultaneity connection and have shown that the
requirement of being a convention strongly restricts its
functional spacetime dependence. In the end we have stated without
proof a theorem which represents a first step towards the search
of simultaneity connections of wider applicability than
Einstein's.


\end{document}